\documentclass[aps,prl,twocolumn,superscriptaddress]{revtex4}

\usepackage{graphicx}

\usepackage{subfigure}

\usepackage{amsmath}

\newcommand{\siesta}[0]{\scriptsize SIESTA \normalsize}
\newcommand{\smeagol}[0]{\scriptsize SMEAGOL \normalsize}

\begin{document}

\title{Control of electron transport through Fano resonances in molecular wires}

\author{T. A. Papadopoulos, I. M. Grace and C. J. Lambert}
\affiliation{Department of Physics, Lancaster University,
Lancaster, United Kingdom}


\begin{abstract}
Using a first principles approach, we study the electron transport
properties of two molecules of length 2.5nm which are the building
blocks for a new class of molecular wires containing fluorenone
units. We show that the presence of side groups attached to these
units leads to Fano resonances close to the Fermi energy. As a
consequence electron transport through the molecule can be
controlled either by chemically modifying the side group, or by
changing the conformation of the side group. This sensitivity,
which is not present in Breit--Wigner resonances, opens up new
possibilities for novel single-molecule sensors.
\end{abstract}

\maketitle

The ability to position single molecules in electrical junctions
\cite{1,2,3,4,5,6} and demonstrations that molecular and nanoscale
structures are capable of basic electronic functions such as
current rectification, negative differential resistance and single
electron transistor behaviour \cite{7,8,9,10,11,12} suggest that
single-molecule electronics may play a key role in the design of
future nanoelectronic circuits. However, the goal of developing a
reliable molecular-electronics technology is still over the
horizon and many key problems, such as device stability,
reproducibility and the control of single-molecule transport need
to be solved. In addition, since contacting molecules via break
junction and SPMs is not a scalable technology, the question of
wiring large numbers of molecules on a single chip still needs to
be addressed.

One approach to developing a scalable technology involves  placing
molecules between arrays of pre-formed, lithographically-grown
contacts. Such an approach requires the length of the molecules to
match the spacing between the contacts and since in practice, the
spacing between such contacts can only be reliably controlled on
the scale of 1--10nm, it is desirable to synthesize families of
molecular wires with this range of lengths.

\begin{figure}
\center{
\includegraphics[scale=0.46]{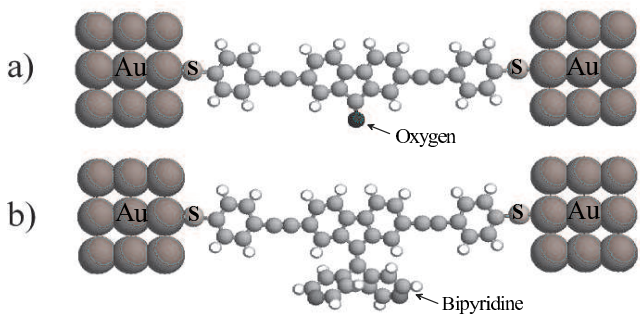}}
\caption{\small The ground state geometry of the 2.5nm molecules
extended to self-consitently include 8 layers of gold on the (111)
surface, each containing 9 Au atoms. Colour codes: C(grey),
H(white), N(dark grey) and O(black). a) The top molecule contains
a single oxygen atom attached to the central fluorenone unit and
b) the oxygen has been replaced by a bi-pyridine
unit.}\label{fig1}
\end{figure}

Recently \cite{Wang1,Wang2,Wang3} a family of rigid molecules has
been synthesized with lengths up to 10nm. These are
$\pi$--conjugated oligomers based on rigid-rod-like
aryleneethynylene backbones, containing fluorenone units. The
presence of terminal thiol groups allow assembly onto gold
surfaces and therefore make them ideal for use in single-molecule
device fabrication. In this Letter we investigate the properties
of the smallest of these molecular wires, of length 2.5nm, since
these form the building blocks of the longer wires. The central
part consists of a single fluorenone unit, which could be
chemically modified, e.g. by replacing the oxygen with pyridine or
bipyridine rings, as shown in Fig.~\ref{fig1}.

In many molecular devices, electron transport is dominated by
conduction through broadened HOMO or LUMO states, leading to
Breit--Wigner resonances \cite{BW}. In contrast, for this family
of molecules, we find that transport is dominated by Fano
resonances \cite{Fano1} associated with the presence of side
groups, such as the oxygen atom or bi-pyridine. Generic properties
of Fano resonances, have been discussed in several contexts
recently. Ref \cite{A} deals with Fano resonances in 1-d
waveguides, applied to GaAS heterojunctions. Ref. \cite{B}
analyses a generic model of a double quantum dot. Refs. \cite{C,D}
address inelastic scattering, which can be an issue at finite
bias, but is not the focus of attention in our paper. Ref.
\cite{E} is concerned with low temperature transport below the
Kondo temperature. 'Conductance cancellation' in 1-d tight-binding
chains has also been noted \cite{F} and a H\"uckel model of
resonant transport recently considered \cite{G}.

None of the above papers contain \emph{ab initio},
material-specific calculations, which treat the metallic
electrodes in a realistic manner. Recently, this limitation has
been removed by state-of-the-art calculations on di-thiol benzene
\cite{H}, which contain many examples of Fano resonances.
Depending on specific conditions, these can take on a wide variety
of shapes, ranging from Breit-Wigner type line shapes to strongly
asymmetric profiles. However the Fano resonances are neither
controlled nor engineered into the molecule. They arise from a
complicated interaction between the molecule and contacts and
cannot be identified with a particular section of the molecule.

The aim of this Letter is to demonstrate that there are advantages
in engineering molecules to possess Fano resonances associated
with specific parts of the molecule, which can be modified
externally to achieve control over transport. Through an \emph{ab
initio} simulation of the molecules in Fig.~1, we demonstrate that
Fano resonances near the Fermi energy can be controlled by
altering the properties of the attached side group. This is in
marked contrast with the behaviour of Breit-Wigner resonances,
which are relatively insensitive to the state of the side group.
These results suggest that the control of Fano resonances opens
intriguing possibilities for single-molecule sensing.

To compute electron transport properties, we use a combination of
the DFT code \siesta \cite{Siesta} and a Green's function
scattering approach, as encapsulated in the molecular electronics
code \smeagol \cite{Smeagol}. Initially the isolated molecule is
relaxed to find the optimum geometry and the molecule is then
extended to include surface layers of the gold leads, so that
charge transfer at the gold-molecule interface is included
self-consistently. The number $N_g$ of incorporated gold layers is
increased until computed transport properties no longer change
with increasing $N_g$. We find that this occurs for $N_g=8$ layers
of gold, each consisting of 9 atoms on the (111) plain. Using a
double--$\zeta$ basis plus polarization orbitals,
Troullier--Martins pseudopotentials \cite{Pseudo} and the
Ceperley--Alder LDA method to describe the exchange correlation
\cite{EC}, an effective tight-binding Hamiltonian of the extended
molecule is obtained, from which a scattering matrix and electron
transmission coefficient $T(E)$ are computed. The zero-bias
electrical conductance is then given by the Landauer formula
$G=(2e^2/h)T(E_F)$, where $E_F$ is the Fermi energy
\cite{Landauer}.

\begin{figure}
\center{\includegraphics[angle=-90,scale=0.3]{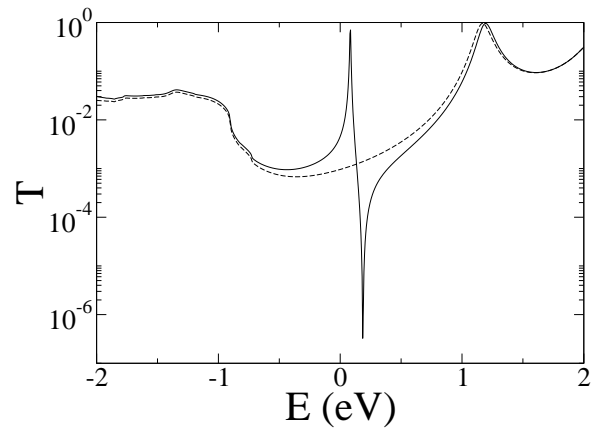}}
\caption{\small Black line: Electron transmission coefficient
versus energy for the molecule having attached an oxygen atom as a
side group. Dashed line: Oxygen bonds removed from the
Hamiltonian.}\label{fig2}
\end{figure}

For the molecule shown in Fig.~\ref{fig1}a, which has an oxygen
atom attached to the fluorenone unit, the transmission coefficient
T(E) is shown in Fig.~\ref{fig2} (black line). For $E \approx
1.2eV$, this exhibits a typical Breit--Wigner resonance. As
expected, since the molecule and contacts are symmetric, the
maximum value of this peak is unity. However in the vicinity of
the Fermi energy ($E_F=0$) transport is dominated by the presence
of an asymmetric Fano resonance \cite{Fano1}, comprising a
resonant peak followed by an anti-resonance.

We now demonstrate that anti-resonances arise from quasi-bound
states associated with the side groups of the molecules in Fig. 1.
For the molecule of Fig. 1a, this is demonstrated by the simple
theoretical trick of artificially setting the hopping matrix
elements connecting the oxygen atom to the fluorenone unit to
zero. This yields the transmission coefficient shown by the red
line in Fig.~\ref{fig2}, which shows that artificial removal of
the chemical bonds to the oxygen destroys the Fano resonance,
whereas the remainder of the transmission spectrum remains largely
unaltered.

\begin{figure}
\center{\includegraphics[angle=-90,scale=0.3]{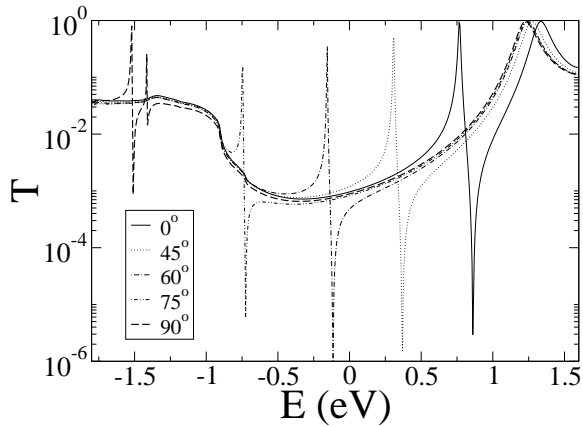}}
\caption{\small Transmission against energy for the bi-pyridine
attached molecule for rotation angles of 0$^\circ$ to
90$^\circ$}\label{fig3}
\end{figure}

The ability to manipulate Fano resonances, by chemical means or
otherwise, opens up new possibilities for controlling
single-molecule transport. To explore this possibility in greater
detail and to demonstrate that Fano resonances are a generic
feature of molecular wires with attached side groups, we examine
the molecule shown in Fig.~\ref{fig1}b, in which the oxygen has
been replaced by a bi-pyridine unit. Recent STM experiments on
related wires have shown that it is possible to change the
rotational conformation of attached side groups \cite{haiss} and
therefore we examine transport properties as a function of the
angle of rotation $\theta$ of the bi-pyridine side group. (We
define the angle $\theta=90^\circ$ when the two rings lie parallel
to the molecule axis and $\theta=0^\circ$ when they lie
perpendicular to it.) The computed transmission through this
molecule is shown in Fig.~\ref{fig3} for five values of $\theta$.
This demonstrates that Fano resonances persist when one side group
(namely the oxygen atom) is replaced by another and furthermore,
the position of the Fano resonance is sensitive to the
conformation of the side group. In contrast, the Breit--Wigner
peak at $1.25eV$ is almost unaffected by such changes.

\begin{figure}
\center{\includegraphics[scale=0.42]{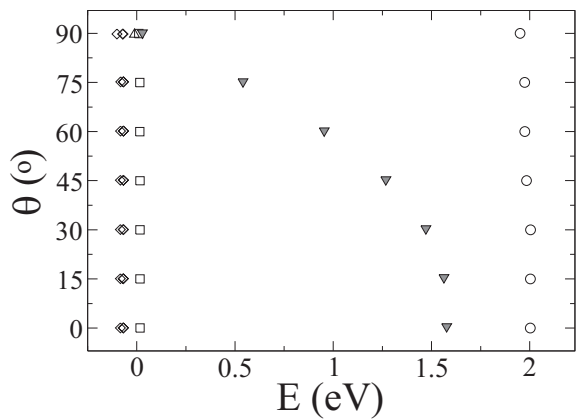}} \caption{\small
Rotation angle $\theta$ dependence of the energy levels of the
isolated molecule. The Fano eigenstates are represented by
triangles while the circles represent the Breit--Wigner energy
levels.}\label{fig4}
\end{figure}

\begin{figure}
\includegraphics[scale=0.43]{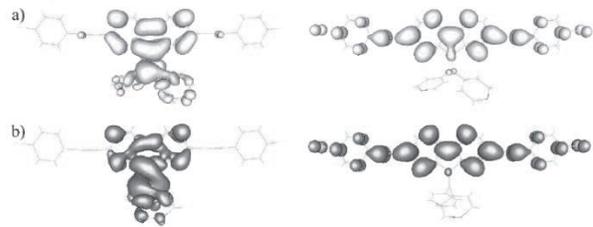}
\caption{\small Constant LDOS surface for a) 30$^\circ$ rotation
and energy values of $1.46eV$ and $2.00eV$. b) 75$^\circ$ rotation
and energy values of $0.54eV$ and $2.00eV$.}\label{fig5}
\end{figure}

To demonstrate that the Fano resonance is associated with
localized states on the side group, we examine the energy spectrum
of the isolated molecule. Fig.~\ref{fig4} shows how the energy
levels of the isolated molecule depend on the rotation angle
$\theta$. This shows that while most of the levels remain
unaffected, one of them is sensitive to changes in $\theta$,
varying from $1.57eV$ at $\theta=90^\circ$ to $0.0eV$ at
$\theta=0^\circ$. To demonstrate that this level belongs to the
fluorenone unit and bi-pyridine, Fig.~\ref{fig5} shows the local
density of states (LDOS) for two different energy values when
$\theta=30^\circ$ (ground state of the molecule) and
$\theta=75^\circ$. Energies $E=2.0eV$ clearly correspond to states
delocalized along the backbone, with almost no weight on the
bi-pyridine. This state is responsible for the Breit--Wigner
resonances in figures 2 and 3. On the other hand, for $E=1.46eV$
and $E=0.54eV$ the orbitals are found to be localized on the
central unit.

\begin{figure}
\center{\includegraphics[scale=0.42]{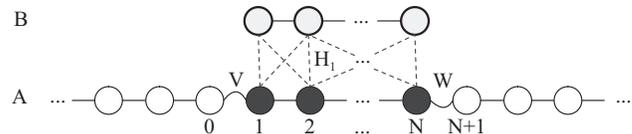}} \caption{\small A
backbone of sites labelled 1 to N coupled to left and right leads
by matrix elements $V, W$ and to a side group of sites by $H_1$.
}\label{fig6}
\end{figure}

To capture the generic features of this combination of a Fano
resonance and nearby Breit--Wigner resonance, we now develop a
model of the system sketched in Fig.~\ref{fig6}, which consists of
a backbone A composed of atomic orbitals numbered $i=1, 2, \ldots,
N$, coupled to a side group B by matrix elements $H_1$. (The
geometric arrangement of the sites is irrelevant, provided A-sites
are weakly coupled to B-sites.) Sites $i \leq 0$ belong to the
left lead and sites $i \ge N+1$ to the right lead. These are
coupled to the ends of the backbone via weak hopping elements $V,
W$.

In the absence of coupling to the side chain, we assume that
transmission through the backbone takes place through a single
backbone state $\Phi_{m}(k)$, with a resonant energy
$\varepsilon_{0}$. In this case, the relevant self-energy $\Sigma$
is

\begin{equation*}
\Sigma=\frac{\omega\omega^{*}}{E-\varepsilon_{0}+i\Gamma}
\end{equation*}
where $\omega=\langle\Psi|H_{1}|\Phi\rangle$. In this expression,
$\Gamma=\Gamma_1+\Gamma_2$ is the broadening due to coupling of
the backbone to the leads, with
$\Gamma_1=|V\Phi_m{(1)}|^{2}N_0(E)$,
$\Gamma_{2}=|W\Phi_m{(N)}|^{2}N_{N+1}(E)$ and $N_0(E)$
($N_{N+1}(E)$) the local density of states for the left (right)
contacts. This yields for the transmission coefficient,

\begin{figure}
\center{
\includegraphics[scale=0.25,angle=-90]{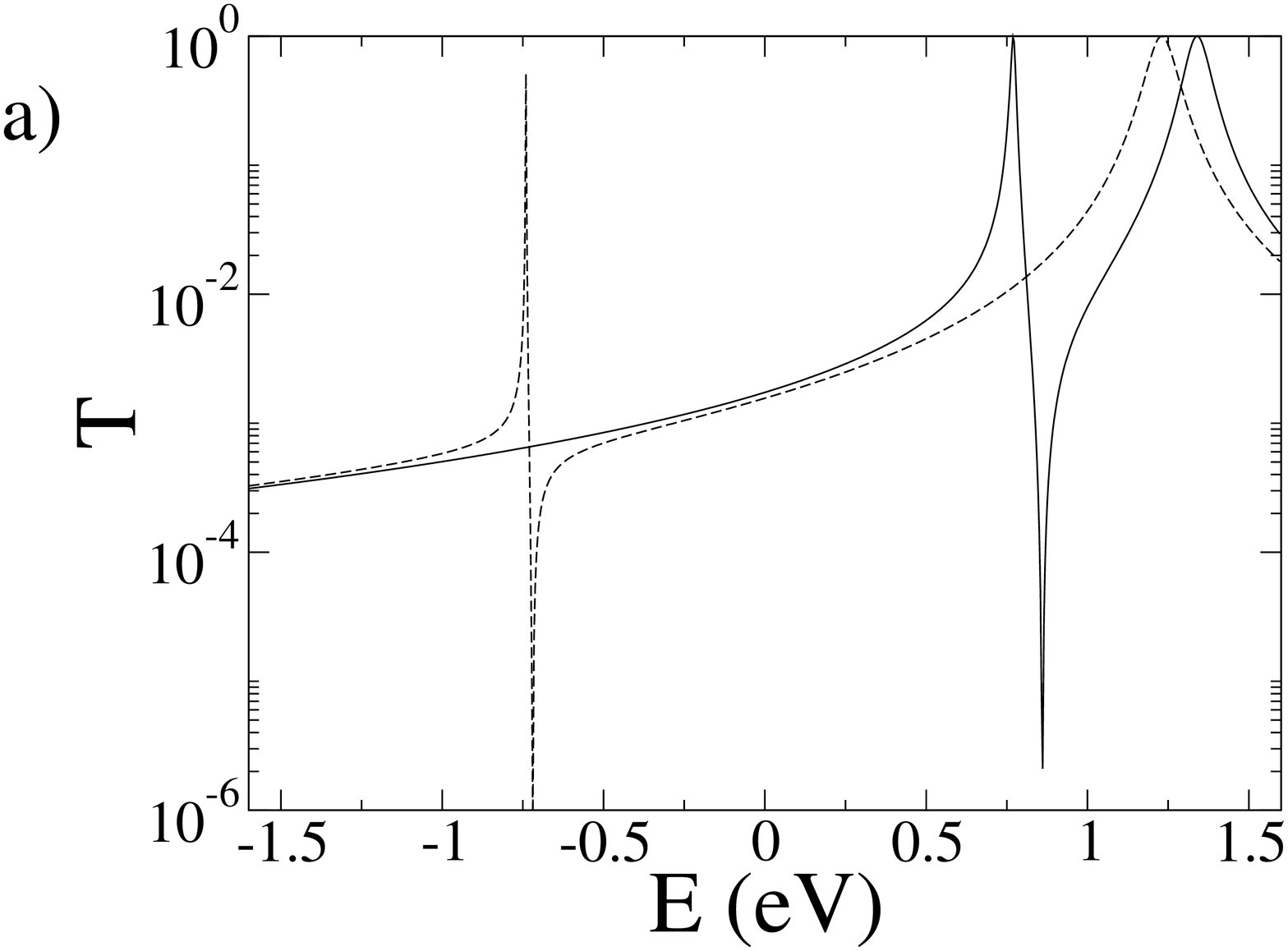}
\includegraphics[scale=0.25,angle=-90]{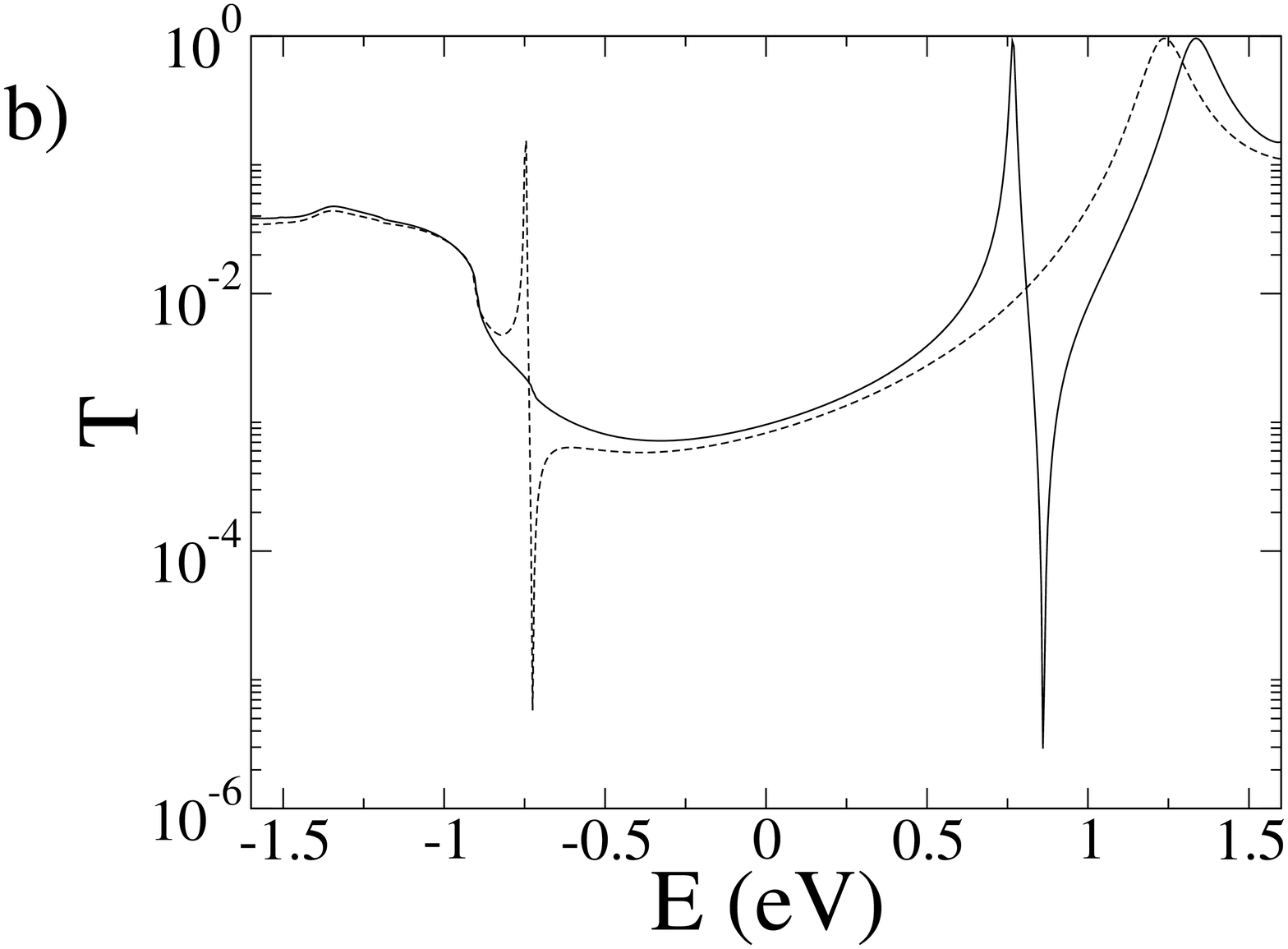}
\caption{\small a) A plot of eq.~(\ref{eq2}) for two sets of
parameters, chosen to fit the \emph{ab initio} results for
$\theta=0^\circ$ and $\theta=75^\circ$. Black line: The values
used are $\varepsilon_{+}=1.34eV$, $\varepsilon=0.86eV$ and
$\varepsilon_{-}=0.77eV$. Dashed line: $\varepsilon_{+}=1.23eV$,
$\varepsilon=-0.72eV$ and $\varepsilon_{-}=-0.74eV$. For both
curves $\Gamma=0.05eV$. b) For comparison, the lower figure shows
the corresponding \emph{ab initio} results.}\label{fig7}}
\end{figure}

\begin{equation}
T\hspace{1mm}=\hspace{1mm}\frac{4\Gamma_{1}\Gamma_{2}}{\left(E-\varepsilon_{0}-
\frac{\omega\omega^{*}}{E-\varepsilon}\right)^2+\Gamma^2}\label{eq2}.
\end{equation}
Equation~(\ref{eq2}) shows that when $\omega=0$, no anti-resonance
occurs and the transmission coefficient exhibits a simple
Breit--Wigner peak. More generally, eq.~(\ref{eq2}) shows that
transmission is a maximum at energies $E=\varepsilon_{\pm}$, where
$\varepsilon_{\pm}$ are the roots of equation
$(E-\varepsilon_{0})(E-\varepsilon)-\omega\omega^{*}=0$ and
vanishes when $E=\varepsilon$. For small $\omega$, a Breit--Wigner
peak of width $\Gamma$ occurs in the vicinity of $\varepsilon_+
\approx \varepsilon_0 $. In addition a Fano peak occurs in the
vicinity of $\varepsilon_{-} \approx \varepsilon$ with width
$\frac{\Gamma\omega\omega^{*}}{(\varepsilon_{0}-\varepsilon)^2}$.
A comparison between eq.~(\ref{eq2}) and the \emph{ab initio}
results of Fig.~\ref{fig3} is shown in figure~\ref{fig7}. This
demonstrates that with an appropriate choice of parameters,
eq.~(\ref{eq2}) captures the essential features of Fano resonances
in aryleneethynylene molecular wires.

In summary we have shown that transport through a new class of
molecular wires, composed of fluorenone subunits with side groups,
is dominated by Fano resonances rather than Breit--Wigner
resonances. As a consequence electron transport through the
molecule can be controlled either by chemically modifying the side
group, or by changing the conformation of the side group. This
sensitivity, which is not present in Breit--Wigner resonances,
opens up the possibility of novel single-molecule sensors.

Acknowledgements: This work was supported by EPSRC under grant
GR/S84064/01 (Controlled Electron Transport) (Durham and
Lancaster), a Lancaster-EPSRC Portfolio Partnership and MCRTN
Fundamentals of Nanoelectronics.


\bibliographystyle{apasoft}

\end{document}